\documentclass[superscriptaddress,aps,a4,prd,twocolumn, nofootinbib]{revtex4-2}
\usepackage{graphicx}

\usepackage[english]{babel}
\usepackage{color}

\usepackage{amsmath,amssymb}
\usepackage{verbatim}
\usepackage[dvipsnames]{xcolor} 

\definecolor{ORANGE}{RGB}{255,165,0}
\colorlet{orange}{ORANGE}
\usepackage{physics}
\usepackage{dcolumn}
\usepackage[a4paper,margin=.7in]{geometry}
\usepackage[normalem]{ulem}
\usepackage{feynmp-auto}
\usepackage{makecell}
\usepackage{amsmath,amssymb}
\usepackage{comment}
\usepackage{slashed}
\usepackage{tikz}
\usetikzlibrary{decorations.markings}
\usepackage{mathtools}

\usepackage{newtxtext, newtxmath}

\let\Im\relax
\DeclareMathOperator{\Im}{Im}
\let\Re\relax
\DeclareMathOperator{\Re}{Re}

\usetikzlibrary{decorations.markings,positioning}

\usepackage[colorlinks=true, allcolors=blue]{hyperref}
\usepackage{booktabs}
\usepackage{cellspace}

\newcommand{\orcid}[1]{\href{https://orcid.org/#1}{	{\includegraphics[height=2ex,width=2ex]{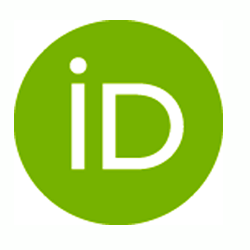}}}}

\begin{document}

\title{A QED Functional Formulation of Relativistic Many-Body Perturbation Theory in Atoms}

  \author{Cedric de Jonge \orcid{0000-0002-1840-6768}} \email{cedric.de-jonge@mpi-hd.mpg.de}
 \affiliation{Max Planck Institute for Nuclear Physics,\\
 Saupfercheckweg 1, 69117 Heidelberg, Germany}

 \author{Zolt\'an Harman \orcid{0009-0006-0554-1537}}
  \affiliation{Max Planck Institute for Nuclear Physics,\\
 Saupfercheckweg 1, 69117 Heidelberg, Germany}
\begin{abstract}
We systematically derive relativistic many-body perturbation theory (RMBPT) from the two-particle irreducible effective action of quantum electrodynamics (QED) in the Furry picture, giving an organisational and diagrammatic framework for both correlation and radiative corrections in many-electron atoms. We further discuss that RMBPT is a well-defined subset of QED effects evaluated on a redefined vacuum, such that the interplay between correlation and radiative effects becomes transparent. We use this to derive QED corrections to the interelectronic interaction of atomic systems, yielding the transverse (frequency-dependent Breit) photon-exchange corrections, including retardation.
Of those corrections, we numerically evaluate the first-order transverse photon-exchange corrections, which complement the conventional RMBPT based on the Dirac-Coulomb no-pair Hamiltonian.
\end{abstract}

\maketitle

\section{Introduction}

Many-body methods \cite{Johnson2007-ip,Sapirstein1998} and quantum electrodynamic (QED) \cite{Shabaev2002, PhysRevLett.85.4699} approaches to atomic systems are historically two separate fields. While the former is mostly concerned with correlation corrections in many-electron systems, the latter is mostly applied to radiative corrections in few-electron systems such as H- to B-like ions. As a consequence, radiative corrections in many-electron atoms are usually a posteriori considerations on top of existing many-body methods. \par
This is an issue that has been partially tackled by Lindgren et al.\ in the past, using their covariant evolution operator formalism in \cite{Lindgren2016,Lindgren2006} to resum interelectronic interaction diagrams with the inclusion of retardation and radiative effects.
Our goal in this work is different in scope.
Rather than attempting to treat QED to all orders, we argue that both approximations can be derived from the QED effective action of the atomic system at hand, as a stationary functional of the electron propagator rather than an energy-dependent effective Hamiltonian on a model space. As a consequence, the Dirac-Hartree-Fock-Coulomb (DHFC) equations and RMBPT can be extracted as well-defined instantaneous, no-pair truncation schemes to this full effective action. Thus, the complement of the truncation, that is, what is left out at every given order of RMBPT, is readily accessible without any additional diagrammatic input. Consequently, this framework makes the interplay of correlation and radiative corrections apparent. It also gives systematic access to corrections that any conventional many-body treatment omits, including the radiative sector. We further note that reformulating the atomic many-body problem within the framework of the quantum effective action opens it up to the plethora of methods developed for such functionals in other fields, such as condensed matter theory, in future work. This has already been done for nonrelativistic density functional theory and the functional renormalization group by Polonyi and Sailer~\cite{PhysRevB.66.155113}. \par
Specifically, we introduce and make use of the two-particle irreducible (2PI) effective action and its relation to the ground-state energy of the system it describes.
The 2PI effective action, introduced by Luttinger and formalized by Cornwall, Jackiw, and Tomboulis~\cite{PhysRevD.10.2428, Luttinger1960}, is widely used in non-equilibrium quantum field theory~\cite{Berges_2002,Aarts_2002,Berges_2004} and condensed matter theory~\cite{Bransch_del_2008,Berges_2007}. For condensate-free theories, its loop expansion coincides with the Luttinger-Ward (LW) functional~\cite{Luttinger1960}, which, together with the closely related variational energy functional of Klein~\cite{PhysRev.121.950}, gives a stationary functional of the propagator and self-energy from which the ground-state energy is obtained. Within this framework, self-consistent approaches to fermion gases are formulated~\cite{Enss_2024,Bauer_2014,Enss_2012}. Moreover, non-relativistic M\o ller-Plesset MBPT has been derived from the LW functional and applied to binding energy calculations in Refs.\ \cite{10.1063/1.1650307,PhysRevA.73.012511,10.1063/1.3089567}. We extend this program to the relativistic QED setting, where the inclusion of radiative corrections becomes natural within the same framework.

Such studies are also motivated by recent applications of atomic structure calculations, one example being the determination of isotope masses for high-precision searches for physics beyond the Standard Model. Isotope-shift measurements in clock transitions are sensitive to  a new force acting between electrons and nucleons because the contribution from the new interaction can depend on the neutron number of the nucleus. For several isotope pairs, a King plot can be constructed, in which a new mediator boson can manifest itself through a nonlinear isotope-shift relation~\cite{Delaunay2017}. An important ingredient in such analyses is the precise knowledge of the isotope masses. In Penning-trap mass-spectrometry experiments, the masses of highly charged ions (HCIs) can nowadays be determined with uncertainties at the eV or even sub-eV level~\cite{Wilzewski2025,Door2025}. Atomic theory is required to account for the mass defect of the electronic binding energies of the corresponding HCI.
Another application is the determination of the neutrino mass through electron capture~\cite{Adam2026,Alpert2025}.
The $Q$ value for this process can be determined from Penning-trap measurements with the mother and daughter ions~\cite{Schweiger2024}.
Here, again, theoretical calculations are required to account for the binding energies of the removed electrons. In both applications,  a clean separation and an accurate treatment of the individual physical contributions is essential for reliable predictions.

In this article, we show that the order-by-order computation of diagrams in the LW skeleton expansion recovers the RMBPT matrix elements and the Dirac-Hartree-Fock-Coulomb (DHFC) equations from a single organisational principle, namely, the QED Luttinger-Ward functional in a Furry-picture reference vacuum.
This provides a functional criterion for separating instantaneous correlation, retardation, negative-energy, self-energy, and vacuum-polarization contributions. The no-pair approximation enters naturally as a splitting of the electron propagator into positive- and negative-energy contributions; pair corrections are then transparent.

We further show that RMBPT is a subset of QED corrections evaluated on the core vacuum:
correlation corrections are precisely the instantaneous Coulomb, no-pair limit of the
self-energy and vacuum polarization diagrams. We show this explicitly through second
order (Secs.~III and IV); for higher-order diagrams it follows by construction. This
identification establishes a systematic scheme for including radiative corrections on
top of correlations and vice versa.
We note that in this work, we follow standard usage and reserve the term radiative corrections for self-energy and vacuum-polarization loops with respect to the QED vacuum, which appear already for hydrogenlike ions and require renormalization. 
Photon exchange between electrons, including its retarded and negative-energy (pair) parts, is referred to as interelectronic interaction or screening effects. 
On the core vacuum, both types of contributions arise from diagrams of self-energy and vacuum-polarization topology.

\section{Theory}

The usual starting point for field-theory based investigations of atomic properties is the QED action in the Furry picture 
\begin{align}
    S_{\text{Furry}} &= \int_x \bar{\psi}(i\slashed{\partial} - m_e + e \slashed{A}_{\text{ext}} + e\slashed{A})\psi -\frac{1}{4}F_{\mu\nu}F^{\mu\nu},
\end{align}
where $\psi$ denotes the electron field, $m_e$ is the electron mass, $e$ stands for the elementary charge, and $\int_x = \int \text{d}^4x$ denotes integration over Minkowski space.
That is, we split the interaction between the electrons and the photon field into a dynamic interaction,
denoted by the four-potential $A$, and an interaction with the external potential $A_{\text{ext}}$, induced by the nucleus. 
The electromagnetic field-strength tensor is defined in terms of the dynamical field as $F_{\mu \nu}=\partial_\mu A_\nu - \partial_\nu A_\mu $.
This separation, in essence, means that we take the nucleus to be infinitely heavy, reducing the interaction with it through photon exchange to an external potential $A_{\text{ext}}$. Of course, this assumption introduces a preferred coordinate system, in which the position of the nucleus sourcing the potential sits at the center. QED in the Furry picture thus breaks translation invariance of the atomic Hamiltonian, and consequently three-momentum is not a conserved quantum number. 
Although obvious, we believe that this is noteworthy in the sense that the usual field theoretical simplifications in momentum space are not straightforwardly applied to the system we consider. In any case, we can still give a generating functional $Z$ for the theory at hand, given by
\begin{align}
    Z[J,\eta,\bar{\eta}] = \int\mathcal{D}\bar{\psi}\mathcal{D}\psi\mathcal{D}A \, e^{i(S_{\text{Furry}} + J\cdot A+\bar{\eta}\cdot \psi + \bar{\psi}\cdot\eta)},
\end{align}
where $J$ is the source field coupled to the photon field, $\eta$ and $\bar{\eta}$ are Grassmann source fields coupled to the Dirac fields $\bar{\psi}$ and $\psi$, respectively, and we use $f\cdot g = \int_x f(x)g(x)$ for brevity\footnote{For generating functionals in a background field, see also Ref.~\cite{lopezlopez2026renormalizedperturbationtheoryintense}.}.
Since we are focused on the interactions between electrons, we integrate out the photon field. To this end, we complete the square with respect to the photon field, resulting in
\begin{align}
    S_{\text{Furry}} &\supset \frac{1}{2}A\cdot D_0^{-1}\cdot A + j_{\psi}\cdot A \nonumber \\
    &= \frac{1}{2}(A+j_{\psi}\cdot D_0)\cdot D^{-1}_0 \cdot(A+D_0 \cdot j_{\psi}) \nonumber \\ &- \frac{1}{2}j_{\psi}\cdot D_0 \cdot j_{\psi}\,, 
\end{align}
with the electron current $j_\psi^\mu = e\bar\psi\gamma^\mu\psi$ and $D_0$ the inverse of the
(gauge-fixed) photon kinetic operator, i.e., the free propagator.
After a field redefinition $A \rightarrow A - (j_{\psi} + J)\cdot D_0$, we can perform the Gaussian integral over photon fields, yielding
\begin{align}
     Z 
     &\sim \int\mathcal{D}\bar{\psi}\mathcal{D}\psi e^{i(S_{\text{eff}}+\bar{\eta}\cdot \psi + \bar{\psi}\cdot\eta)},
\end{align}
where the new effective action is given by the Furry picture including a photon mediated bilocal four-point interaction
\begin{align}
    S_{\text{eff}} &= \int_x \bar{\psi}(i\slashed{\partial} - m_e + e \slashed{A}_{\text{ext}})\psi\nonumber \\
    &- \frac{e^2}{2} \int_{x,y} \bar{\psi}(x)\gamma^{\mu}\psi(x) D_{\mu\nu}(x-y)\bar{\psi}(y)\gamma^{\nu}\psi(y).
\end{align}
In the Coulomb gauge with the instantaneous approximation
$D_{00}(x-y)\to V(\mathbf{x}-\mathbf{y})\delta(x^0-y^0)$,
$D_{ij}=0$, the interaction in the second row of the above equation reduces to
\begin{equation}
  -\frac{e^2}{2}\int_{x,y}
    \psi^\dagger(x)\psi(x)\,V(\mathbf{x}-\mathbf{y})\,
    \psi^\dagger(y)\psi(y)\,\delta(x^0-y^0),
\label{eq:Coulomb_limit}
\end{equation}
which we use whenever we consider the instantaneous limit.
Now, we also modify the generating functional to include a bilocal source $K(x,y)$ and set external photon sources to zero 
\begin{align}
    Z[K,\bar{\eta},\eta] = \int \mathcal{D}\bar{\psi}\mathcal{D}\psi\, e^{i(S_{\text{eff}}+ \bar{\psi}\cdot K\cdot \psi  +\bar{\eta}\cdot \psi + \bar{\psi}\cdot\eta)}.
\end{align}
The 2PI effective action is then defined as the double Legendre transform
\begin{align}
    \Gamma[G,\pmb{\Psi}] &= -i\ln Z[K,\pmb{\eta}] - K\cdot G 
    - \bar{\eta}\cdot \Psi -\bar{\Psi}\cdot\eta \nonumber \\
    &-\bar{\Psi}\cdot K \cdot \Psi,
\end{align}
where $\Psi = \langle \psi\rangle$, and $G = \langle \mathcal{T}\psi\bar{\psi}\rangle$. Since we consider atomic systems without an electron condensate, we set $\Psi =0 $. The effective action then admits a decomposition into a free part and the \textit{Luttinger-Ward} functional $\Phi$~\cite{Luttinger1960}, that is, 
\begin{align}
    \Gamma[G] = \Gamma_0[G] -i \Phi[G].
\end{align}
The zeroth order contribution to the effective action is 
\begin{align}
    \Gamma_0[G] =-i\text{Tr}\ln G^{-1} - i\text{Tr} G_0^{-1}G.
\end{align}
It is instructive to consider the stationarity condition at first order, which is given by 
\begin{align}
    \frac{\delta \Gamma_0[G]}{\delta G}\Bigg|_{K=0} = 
    i(G^{-1} - G_0^{-1}) &= 0 , 
\end{align}
meaning that the propagator is given by the tree-level solution at zeroth order\footnote{We want to emphasize at this point that this only applies in the limit $\Psi = 0 $, otherwise this propagator would correspond to the nonlinear equations of motion.}.
The Luttinger-Ward functional encodes all 2PI corrections that are given in a skeleton expansion
\begin{align}
\label{eq:skeleton}
\Phi[G] &= \quad
\vcenter{\hbox{
\begin{fmffile}{Hartree}
\begin{fmfgraph*}(10,30)  
  \fmfforce{(0.1w,0.8h)}{v1}
  \fmfforce{(0.1w,0.2h)}{v2}
  \fmf{double,left=0.8,tension=1}{v1,v1}
  \fmf{double,left=0.8,tension=1}{v2,v2}
  \fmf{photon}{v1,v2}
  \fmfv{decor.shape=circle,decor.filled=full,decor.size=thick}{v1,v2}
\end{fmfgraph*}  
\end{fmffile}
}}
+
\vcenter{\hbox{
\begin{fmffile}{Fock}
\begin{fmfgraph*}(50,50)
  \fmfset{arrow_len}{0mm}
  \fmfforce{(0.5w,0.8h)}{v1}
  \fmfforce{(0.5w,0.2h)}{v2}
  \fmf{double,left=0.8,tension=1}{v1,v2}
  \fmf{double,left=0.8,tension=0.1}{v2,v1}
  \fmf{photon}{v1,v2}
  \fmfv{decor.shape=circle,decor.filled=full,decor.size=thick}{v1,v2}
\end{fmfgraph*}  
\end{fmffile}
}}
+
\vcenter{\hbox{
\begin{fmffile}{Second-Order}
\begin{fmfgraph*}(60,70)
  \fmfset{arrow_len}{0mm}
  \fmfforce{(0.2w,0.65h)}{v1}
  \fmfforce{(0.2w,0.35h)}{v2}
  \fmfforce{(0.8w,0.65h)}{v3}
  \fmfforce{(0.8w,0.35h)}{v4}
  \fmf{double,left=0.2,tension=1}{v1,v3}
  \fmf{double,left=0.2,tension=1}{v3,v1}
  \fmf{double,left=0.2,tension=1}{v4,v2}
  \fmf{double,left=0.2,tension=1}{v2,v4}
  \fmf{photon}{v1,v2}
  \fmf{photon}{v3,v4}
  \fmfv{decor.shape=circle,decor.filled=full,decor.size=thick}{v1,v2,v3,v4}
\end{fmfgraph*} 
\end{fmffile}
}}
\nonumber \\[-2em] &+
\vcenter{\hbox{
\begin{fmffile}{Second-Order-Exch}
\begin{fmfgraph*}(60,70)
  \fmfset{arrow_len}{0mm}
  \fmfforce{(0.2w,0.65h)}{v1}
  \fmfforce{(0.2w,0.35h)}{v2}
  \fmfforce{(0.8w,0.65h)}{v3}
  \fmfforce{(0.8w,0.35h)}{v4}
  \fmf{double,left=0.2,tension=1}{v1,v3}
  \fmf{double,tension=1}{v3,v2}
  \fmf{double,tension=1}{v4,v1}
  \fmf{double,left=0.2,tension=1}{v4,v2}
  \fmf{photon}{v1,v2}
  \fmf{photon}{v3,v4}
  \fmfv{decor.shape=circle,decor.filled=full,decor.size=thick}{v1,v2,v3,v4}
\end{fmfgraph*} 
\end{fmffile}
}}
+ \quad \dots,
\end{align}
where the first two diagrams can be identified with the Dirac-Hartree-Fock approximation, i.e., the first-order result in the RMBPT framework. The third and the fourth term can be identified as the second-order correlation correction. In total, we can identify the functional derivative of said object as the self-energy, meaning that the stationarity condition yields the Dyson-Schwinger equation (DSE)
\begin{align}
    G^{-1} = G_0^{-1} +\Phi'[G]. \label{eq:DSE}
\end{align}
The effective action evaluated at the physical propagator gives direct access to the ground-state energy of the theory. To see this, recall that the partition function projects onto the ground state in the limit $T\to(1-i\epsilon)\infty$,
\begin{align}
Z = \langle\Omega|e^{-iHT}|\Omega\rangle &= \sum_n |\langle\Omega|n\rangle|^2 e^{-iE_nT} \nonumber \\
&\xrightarrow{T\to(1-i\epsilon)\infty} |\langle\Omega|0\rangle|^2 e^{-iE_0 T},
\end{align}
since the oscillating contributions from excited states are suppressed by the $i\epsilon$ prescription. Taking the logarithm and dividing by $iT$, and using $\Gamma \sim  i\ln Z$ at the saddle point $K=0$, $\pmb{\eta}=0$:
\begin{align}
\lim_{T\to(1-i\epsilon)\infty}\frac{\Gamma}{
T} = -E_0 + \text{const},
\end{align}
where the additional constant is the infinite vacuum energy, meaning we need a suitable reference.
The higher-order corrections from the Luttinger-Ward functional therefore directly yield corrections to $E_0$, order by order in the skeleton expansion. In the following we write $T = \int_{-\infty}^\infty dt = 2\pi \delta(0)$, understanding that the infinite time integral is factored out in computing the energy corrections.

\subsection{Choosing the Reference Vacuum \& The No-Pair approximation}
We take the common choice of vacuum for the theory, and define it to be the occupied electron core of the atom in question \cite{sym13061014} (while neglecting negative energy contributions)
\begin{align}
    \ket{\Omega} = \prod_{a\in \text{core}} d^{\dagger}_a \ket{0},
\end{align}
with $d^{\dagger}_a$ being the creation operator for an electron in a bound Dirac-Fock atomic state $a$, and $\ket{0}$ the unoccupied QED vacuum.
Most notably, the occupied vacuum enters the definition of the propagator as 
\begin{align}
    G_{\alpha\beta}(x,y) = \bra{\Omega} \mathcal{T}\{\psi_{\alpha}(x)\bar{\psi}_{\beta}(y)\}\ket{\Omega} \,.
\end{align}
The time ordering $\mathcal{T}$ then naturally splits the no-pair propagator into core and excited states as 
\begin{align}
    G^+(x,y) &= \frac{i}{2\pi}\int\text{d}\omega \,e^{-i\omega(x^0-y^0)}\Bigg[\sum_{a\in \text{core}}\frac{\psi_a(\pmb{x})\bar{\psi}_a(\pmb{y})}{\omega-(\epsilon_a+i\gamma)} \nonumber \\[.5em]
    &+ \sum_{n \in \text{excited}} \frac{\psi_n(\pmb{x})\bar{\psi}_n(\pmb{y})}{\omega-(\epsilon_n-i\gamma)}\Bigg], \nonumber \\[.5em]
    G^{-}(x,y) & = \frac{i}{2\pi}\int\text{d}\omega\, e^{-i\omega(x^0-y^0)}\sum_{\epsilon_n < 0 } \frac{\psi_n(\pmb{x})\bar{\psi}_n(\pmb{y})}{\omega-(\epsilon_n+i\gamma)}.
\end{align}
where the sum over excited states in the second row includes an implicit integration over the positive-energy continuum. 
We further define the propagator on the QED vacuum as $\tilde{G}$, i.e., 
\begin{align}
    \tilde{G}_{\alpha\beta}(x,y) = \bra{0} \mathcal{T}\{\psi_{\alpha}(x)\bar{\psi}_{\beta}(y)\}\ket{0} \,.
\end{align}
Recalling the need for a reference to extract energies from the effective action, we choose the unoccupied vacuum as said reference, i.e., we define ground-state energy as the difference w.r.t.\ the ground-state energy of the theory defined on the unoccupied vacuum
\begin{align}
    \Delta \Gamma[G] = \Gamma[G]-\Gamma[\tilde{G}],
\end{align}
and compute the ground state energy following the variational construction of Ref.\ \cite{PhysRev.121.950}
\begin{align}
    E_0 = -\frac{\Delta \Gamma[G]}{T}
\end{align}
We will see that this choice recovers the no-pair approximation as a subset of diagrams, as we will show explicitly for the first order case in the following section. We further introduce the following diagrammatic notation for different propagators
\begin{align}
    G &= 
    \begin{gathered}
\raisebox{10pt}{
    \begin{fmffile}{G}
        \begin{fmfgraph*}(50,10)
        \fmfleft{i1,i2}
        \fmfright{o1,o2}
        \fmf{double}{i1,o1}
        \end{fmfgraph*}
    \end{fmffile}}
\end{gathered}\nonumber \\
G^+ &= 
\begin{gathered}
\raisebox{10pt}{
    \begin{fmffile}{Gp}
        \begin{fmfgraph*}(50,10)
        \fmfleft{i1,i2}
        \fmfright{o1,o2}
        \fmf{double,label=$+$,l.d=1pt}{i1,o1}
        \end{fmfgraph*}
    \end{fmffile}}
\end{gathered} \nonumber \\
\end{align}
We do not introduce a diagrammatic expression for $G^-$, since it will not appear in a diagram in this work.
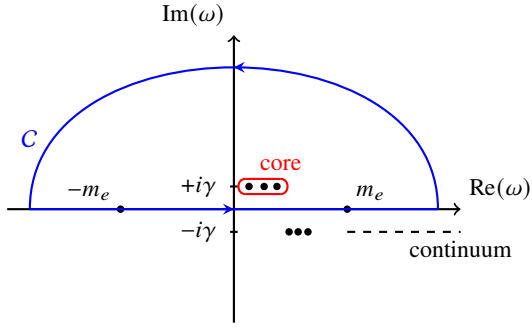
\begin{figure}
    \centering
\begin{tikzpicture}[thick]

  \draw [->] (-3,0) -- (3,0) node [above
  right] {$\Re(\omega)$};
  \draw [->] (0,-1.5) -- (0,2.3) coordinate [below left = 0.3 and 0.1] (y-axis) node [above left] {$\Im(\omega)$};

  \filldraw[black] (0.2,.3) circle (1pt);
  \filldraw[black] (0.4,.3) circle (1pt);
  \filldraw[black] (0.57,.3) circle (1pt);
  \filldraw[black] (0.73,-.3) circle (1pt);
  \filldraw[black] (0.85,-.3) circle (1pt);
  \filldraw[black] (0.98,-.3) circle (1pt);
  \filldraw[black] (-1.5,0) circle (1pt) node [above left] {$-m_e$};
  \filldraw[black] (1.5,0) circle (1pt) node [above right] {$m_e$};

  \draw[blue,thick,postaction={decorate}, 
      decoration={markings, mark=at position 0.5 with {\arrow{stealth}}}] (2.7,0) .. controls +(up:2.5) and +(up:2.5) ..(-2.7,0) node [above =0.7cm] {$\mathcal{C}$};
  \draw[blue,thick,postaction={decorate}, 
      decoration={markings, mark=at position 1 with {\arrow{stealth}}}] (-2.7,0) -- (-0,0);
  \draw[blue,thick] (2.7,0) -- (0,0);

  \draw[-] (-.05,-.3) -- (.05,-.3) node [left] {$-i\gamma \,\,\,$};
  \draw[-] (-.05,.3) -- (.05,.3) node [left] {$+i\gamma \,\,\,$};
  \draw[dashed] (1.5,-.3) -- (3,-.3) node [below] {continuum};
  \draw[red] (0.15,.4) --  (0.62,.4) node [above] {core};
  \draw[red] (0.15,.2) --  (0.62,.2);
  \draw[red] (0.15,.2) .. controls +(left:.12) and +(left:.12)..(0.15,.4);
  \draw[red] (0.62,.2) .. controls +(right:.12) and +(right:.12)..(0.62,.4);
\end{tikzpicture}
\caption{Schematic of the frequency integration for fermionic vacuum bubbles in the no-pair approximation. The latter expresses itself in the missing branch cut starting at $-m_e$. Closing in the upper half plane means that we only pick up the vacuum contributions, which in this case are the occupied core orbitals.}
\end{figure}

\subsection{Notation}
In the following, we will use consistent notation with regard to the labelling of states to ensure that the expressions are more easily understood. The assignments are 
\begin{itemize}
    \item $a,b,c,d$ denote core states
    \item $m,n$ denote excited states, that is, positive-energy states that are not core states
    \item $v$ denotes valence states
    \item $k,l$ denote sea, or negative energy states
\end{itemize}
unless otherwise specified within this paper. Using the tools shown above, we can then proceed to evaluating energy shifts within the abovementioned skeleton expansion of the 2PI effective action.

\subsection{The First-Order Diagrams}

For simplicity, we consider the Coulomb interaction $D_{00}(x-y) = V(x-y)$ separately, denoting the resulting \textit{instantaneous} LW functional as $\tilde{\Phi}$, and the remainder, i.e, the diagrams containing only the retarded photon, as $\bar{\Phi} = \Phi - \tilde{\Phi}$. This split naturally arises when considering the photon propagator in the Coulomb gauge, which is the natural choice for RMBPT or other methods for extracting Coulomb correlation corrections to atomic systems. The resulting $\tilde{\Phi}$ diagrams however, can easily be extended to also include instantaneous Breit interactions as well. The first diagram in the skeleton expansion of Eq.~(\ref{eq:skeleton}) is then given by 
\begin{align}
    \frac{-ie^2}{2}\int_{x,y} V(x-y) \text{Tr}(\gamma^0G(x,x))\text{Tr}(\gamma^0G(y,y))
\end{align}
Considering the propagator by itself, we obtain 
\begin{align}
    G(x,x) = \frac{i}{2\pi} \int \text{d}\omega \, G(\pmb{x},\pmb{x},\omega) e^{-i\omega 0^-}.
\end{align}
Here and in what follows, a bold symbol denotes the spatial component of a fourvector.
We perform the frequency integration in the upper complex half plane, such that we pick up the occupied (core) states and negative continuum only, i.e., 
\begin{align}
    G(x,x) = -\sum_{a \in \text{core}} \psi_a(\pmb{x})\bar{\psi}_a(\pmb{x}) - \sum_{\epsilon_n < 0 } \psi_n(\pmb{x})\bar{\psi}_n(\pmb{x}).
\end{align}
Similarly, we have for the same diagram on the QED vacuum 
\begin{align}
     \tilde{G}(x,x) = -\sum_{\epsilon_n < 0 } \psi_n(\pmb{x})\bar{\psi}_n(\pmb{x})
\end{align}
meaning that the full first order direct diagram including the reference is given by 
\begin{align}
    \tilde{\Phi}^{(1)}_{\text{dir}}[G] &- \tilde{\Phi}^{(1)}_{\text{dir}}[\tilde{G}] \nonumber \\
    &= -\frac{i e^2 T}{2} \sum_{a,b\in \text{core}} \int_{\pmb{x},\pmb{y}} V(\pmb{x}-\pmb{y})  \rho_{aa}(\pmb{x})\rho_{bb}(\pmb{y}) \nonumber \\[1em]
    &-  i e^2 T \sum_{\substack{\epsilon_n < 0 \\ b \in \text{core}}} \int_{\pmb{x},\pmb{y}} V(\pmb{x}-\pmb{y}) \rho_{nn}(\pmb{x})\rho_{bb}(\pmb{y}),
\end{align}
where we defined $\rho_{aa}(\pmb{x}) = \psi^{\dagger}_a(\pmb{x})\psi_a(\pmb{x})$.
We can now quite straightforwardly identify different parts of this skeleton with known quantities. The first term is the Hartree term associated with DHFC in the no-pair approximation, while the latter is a vacuum polarization-type contribution. Specifically, this means that
\begin{align}
    \tilde{\Phi}^{(1)}_{\text{dir}}[G] - \tilde{\Phi}^{(1)}_{\text{dir}}[\tilde{G}] = \tilde{\Phi}^{(1)}_{\text{Hartree}} + \tilde{\Phi}^{(1)}_{\text{VP}}
\end{align}
The second diagram in the expansion shown in Eq.~(\ref{eq:skeleton}) is given by the expression  
\begin{align}
    \frac{ie^2}{2}\int_{x,y} V(x-y) \text{Tr}(\gamma^0 G(x,y)\gamma^0G(y,x)).
\end{align}
Considering that $V$ is an instantaneous interaction, we get a factor of $\delta(x^0-y^0)$, meaning the frequency integration is once again evaluated at equal times
\begin{align}
    -\frac{ie^2}{2(2\pi)^2}\int_{\omega,\omega'} G(\pmb{x},\pmb{y},\omega) G(\pmb{y},\pmb{x},\omega') e^{-i\omega 0^-+ i \omega' 0^+}.
\end{align}
This means that the frequency integrals decouple, leaving us with only the residues associated with the poles in the upper half plane, that is, 
\begin{align}
    \tilde{\Phi}^{(1)}_{\text{exch}}[G] &-  \tilde{\Phi}^{(1)}_{\text{exch}}[\tilde{G}]\nonumber \\ 
    &= \frac{ie^2 T}{2}\sum_{a,b\in \text{core}} \int_{\pmb{x},\pmb{y}} V(\pmb{x}-\pmb{y}) \rho_{ab}(\pmb{x})\rho_{ba}(\pmb{y})\nonumber \\
    &+ie^2 T\sum_{\substack{a\in\text{core}\\\epsilon_n < 0}} \int_{\pmb{x},\pmb{y}} V(\pmb{x}-\pmb{y})  \rho_{an}(\pmb{x})\rho_{na}(\pmb{y}),
\end{align}
where we defined $\rho_{ab}(\pmb{x}) = \psi_{a}^{\dagger}(\pmb{x})\psi_b(\pmb{x})$.
Here, the diagrams again decompose into a Fock, and a SE-type contribution, meaning that 
\begin{align}
    \tilde{\Phi}^{(1)}_{\text{exch}}[G] -  \tilde{\Phi}^{(1)}_{\text{exch}}[\tilde{G}]= \tilde{\Phi}^{(1)}_{\text{Fock}} + \tilde{\Phi}_{\text{SE}}.
\end{align}
Reintroducing the retarded part of the skeleton $\bar{\Phi}^{(1)}$ then, in a similar fashion, introduces retardation corrections to screening, as well as the retarded parts of the first order radiative corrections associated with self energy and vacuum polarization diagrams. To show this, we use the Sokhotski-Plemelj theorem
\begin{align}
    \frac{1}{x-i0} - \frac{1}{x+i0} = 2\pi i \delta(x),
\end{align}
which collapses the difference of the propagator on the redefined vacuum and the empty vacuum onto a sum of core states, i.e., 
\begin{align}
    G(x,y) - \tilde{G}(x,y) = - \sum_{a\in\text{core}} e^{-i\epsilon_a(x^0-y^0)}\psi_a(\pmb{x})\bar{\psi}_a(\pmb{y}),
\end{align}
so rewriting $G = \tilde{G} + G-\tilde{G}$ then yields cutting rules, which applied to the retarded direct diagram results in
\begin{widetext}
\begin{align}
    &\bar{\Phi}_{\text{dir}}^{(1)}[G] - \bar{\Phi}_{\text{dir}}^{(1)}  [\tilde{G}]\nonumber \\
    &= -\frac{ie^2T}{2} \int_{x,y} D_{ij}(x-y) \Bigg[ \sum_{a,b}(\bar{\psi}_a(\pmb{y})\gamma^i\psi_a(\pmb{y}))(\bar{\psi}_b(\pmb{x})\gamma^j\psi_b(\pmb{x}))e^{-i(\epsilon_a+\epsilon_b)0^+}-2\text{Tr}(\gamma^i\tilde{G}(x,x))\sum_{a}\bar{\psi}_a(\pmb{y})\gamma^j\psi_a(\pmb{y})e^{-i\epsilon_a0^+}\Bigg].
    \label{eq:retardation_dir}
\end{align}
\end{widetext}
The first contribution is to be associated with the retardation correction to the no-pair part of the instantaneous diagram, whereas the other is a genuine radiative correction that we group with the VP-type diagram we saw earlier. 
An analogous treatment will give the corresponding corrections for the exchange contribution, i.e., self-energy and exchange-type retarded screening. 

\subsection{The Dirac-Hartree-Fock Equations}

Using these results, we can consider the Dyson-Schwinger equation at first order in the skeleton expansion, which is given by Eq.\ (\ref{eq:DSE}) taken up to the first order in the (instantaneous) skeleton expansion
\begin{align}
    G_{\text{DHFC}}^{-1}(x,y) = G^{-1}_{0}(x,y) +\tilde{\Phi}^{'(1)}[G^+].
\end{align}
Essentially, we are opening up lines in the diagrammatic expansion and re-expressing the corrections as self-energy insertions. This we can then use to obtain the modified Dirac equation.
The corresponding derivatives are 
\begin{align}
   \tilde{\Phi}_{\text{dir}}^{'(1)}[G](z,z') &= ie^2\gamma^0\int_{\pmb{x}}V(\pmb{z}-\pmb{x}) \Bigg[\sum_{a\in \text{core}}\psi^{\dagger}_a(\pmb{x})\psi_a(\pmb{x})\nonumber \\
   &+\sum_{\epsilon_k <0} \psi^{\dagger}_{k}(\pmb{x})\psi_{k}(\pmb{x})\Bigg]\delta(z-z')\nonumber \\
   \bar{\Phi}_{\text{dir}}^{'(1)}[G](z,z') &= -ie^2 \int_xD_{ij}(x-z) \text{Tr}(\gamma^iG(x,x))\gamma^j\nonumber \\
   &\times \delta(z-z')
\end{align}
which gives a clean separation into instantanous screening, retarded screening, and vacuum polarization. Similarly, we obtain for the first order exchange skeleton 
\begin{align}
    \tilde{\Phi}^{'(1)}_{\text{exch}}[G](\pmb{z},\pmb{z}') &= -ie^2 \gamma^0V(\pmb{z}-\pmb{z'})\Bigg[\sum_{a \in \text{core}}
    \psi_{a}(\pmb{z})\psi^{\dagger}_{a}(\pmb{z}')\nonumber \\
    &+ \sum_{\epsilon_k <0} \psi_{k}(\pmb{z})\psi^{\dagger}_{k}(\pmb{z}')\Bigg]\nonumber \\  
    \bar{\Phi}^{'(1)}_{\text{exch}}[G](z,z') &=ie^2  D_{ij}(z-z')\gamma^i G(z,z')\gamma^j.\nonumber \\
\end{align}
The instantaneous part of both diagrams can be split into two distinct contributions, a radiative one, containing the negative energy continuum, and the no-pair one, which is the one we will further use for the DSE. 
Now, consider the inverse zeroth order propagator, which can be given in the form dictated by the equations of motion 
\begin{align}
    -i(i\slashed{\partial}-m_e+e\gamma^0V(|\pmb{x}|)) 
    &=-i(\gamma^0\omega - \gamma^0 h_{\rm D}),
\end{align}
where $h_{\rm D}$ is the Dirac-Coulomb Hamiltonian. This allows us to rewrite the DSE as a modified Dirac Coulomb equation. We arrive there by noting that 
\begin{align}
    \delta(x-y) &= \int_aG_0^{-1}(x-a)G(a-y) \nonumber \\
    &+ \int_a\tilde{\Phi}^{'(1)}[G^+](x-a)G(a-y),
\end{align}
and inserting the spectral representations of the operators. Since the instantaneous $\tilde{\Phi}^{'(1)}$ is frequency independent, this yields 
\begin{align}
    \delta(\pmb{x}-\pmb{y}) &= \sum_n
    \frac{1}{\omega-\epsilon_n\pm i\gamma} \nonumber \\
    &\times\Bigg[ \gamma^0(\omega-h_{\rm D})\psi_n(\pmb{x})
    \nonumber \\
    &+ i\int_{\pmb{a}}\tilde{\Phi}^{'(1)}[G^+](\pmb{x},\pmb{a})\psi_n(\pmb{a})\Bigg]
    \bar{\psi}_n(\pmb{y}).
\end{align}
    The LHS of this equation has no poles, meaning that the residue at $\omega=\epsilon_i$ must vanish for every $\pmb{y}$. Multiplying by $(\omega-\epsilon_i)$, taking the limit $\omega\rightarrow\epsilon_i$, and multiplying by $\gamma^0$ from the left then yields
\begin{align}
    (\epsilon_i-h_{\rm D})\psi_i(\pmb{x}) + i\gamma^0\int_y \tilde{\Phi}^{'(1)}[G^+](\pmb{x},\pmb{y})\psi_i(\pmb{y}) = 0 \,.
\end{align}
which means that in total, we obtain the recursive Dirac-Hartree-Fock equations
\begin{widetext}
\begin{align}
\label{eq:dhf}
    (\epsilon_i-&h_{\rm D})\psi_i(\pmb{x}) - e^2\sum_{a\in\text{core}}\bigg[\int_{\pmb{z}}\psi^{\dagger}_a(\pmb{z})\psi_{a}(\pmb{z}) V(\pmb{x}-\pmb{z})\psi_i(\pmb{x}) - \int_{\pmb{y}} V(\pmb{x}-\pmb{y}) \psi_a(\pmb{x})\psi^{\dagger}_a(\pmb{y}) \psi_i(\pmb{y}) \bigg]=0,
\end{align}
\end{widetext}
which is in agreement with previous derivations of the DHFC equations (see e.g.~\cite{Swirles1935,Johnson2007-ip,Grant2023}). 
Now, we want to make a point regarding the diagrammatic interpretation and origin of these equations. We essentially derived them from the first-order instantaneous DSE on the non-trivial core vacuum, meaning the solution of the self consistent equations of motion are, in essence, a resummation of the corresponding DSE. They can be given by the following diagrammatic expression
\begin{align}
    G_{\text{DHFC}}^{-1}
    &= \quad G_0^{-1} \quad + \quad 
   \begin{gathered}
    \begin{fmffile}{DSE-Full_1}
    \begin{fmfgraph*}(25,30)
    \fmfleft{i1}
    \fmfright{o1}
    \fmfforce{(0w,0.25h)}{i1}
    \fmfforce{(1w,0.25h)}{o1}
    \fmf{double}{i1,v1,o1}
    \fmffreeze
    \fmfforce{(0.5w,0.25h)}{v1}
    \fmfforce{(0.5w,0.7h)}{v2}
    \fmf{dashes}{v1,v2}
    \fmf{double, label=$+$,l.d=1pt,left=0.8,tension=1}{v2,v2}
    \fmfv{decor.shape=circle,decor.filled=full,decor.size=1.4thick}{v1,v2}
    \end{fmfgraph*} 
    \end{fmffile}
\end{gathered}
\quad
+ 
\begin{gathered}
\raisebox{13pt}{
    \begin{fmffile}{DSE-Full_2}
        \begin{fmfgraph*}(50,30)
        \fmfleft{i1,i2}
        \fmfright{o1,o2}
        \fmf{phantom}{i1,o2}
        \fmf{double,tension=.5}{i1,v11}
        \fmf{double,tension=.5}{o1,v12}
        \fmf{double, label=$+$,l.d=1pt,tension=.1}{v11,v12}
        \fmf{dashes,tension=0,left=1}{v11,v12}
        \fmfv{decor.shape=circle,decor.filled=full,decor.size=1.4thick}{v11,v12}
        \end{fmfgraph*}
    \end{fmffile}}
\end{gathered}
\end{align}
where the dashed lines indicate instantaneous Coulomb propagators and the plus signs indicate no-pair propagators. From this, it is clear that employing self consistent field methods amounts to a partial resummation of the QED self energy. The missing parts are the negative energy continuum, as well as the frequency-dependent (transverse) contributions shown in the previous equations. \par
We thus conclude that a consistent treatment of self-energy in conjunction with self consistent field method requires careful bookkeeping of the terms already included in the resummation scheme. This is akin to the treatment of relativistic photon exchange corrections in the extended Furry picture \cite{PhysRevA.107.032815}. The framework of this paper makes this bookkeeping a natural consequence of the derivation.

\subsection{Retardation Corrections to the DHFC Equations}

On top of the previously shown resummation of instantaneous interactions, one can consider retardation corrections to this first order. We account for those retardation effects perturbatively, which in the case of core-correction simply means computing the subset of first order skeletons not contained in the DHFC resummation scheme. Namely, this means computing the shift induced by Eq.\ (\ref{eq:retardation_dir}). In this work, we focus on the transverse (Breit) photon-exchange corrections to the DHFC energy, i.e., 
\begin{align}
    &\frac{e^2}{2T} \int_{x,y}D_{ij}(x-y)  \sum_{a,b}\text{Tr}(\gamma^i\psi_a(\pmb{y})\bar{\psi}_a(\pmb{y}))\nonumber \\
    &\times\text{Tr}(\gamma^j\psi_b(\pmb{x})\bar{\psi}_b(\pmb{x}))\nonumber \\
    &= \frac{e^2}{2} \sum_{a,b} \int_{\pmb{x},\pmb{y}} \psi^{\dagger}_a(\pmb{x})\psi^{\dagger}_b(\pmb{y}) \alpha^i\alpha^j D_{ij}(\pmb{x}-\pmb{y};0) \psi^{}_a(\pmb{x})\psi^{}_b(\pmb{y}),
\end{align}
with $\alpha^i =\gamma^0\gamma^i$. Next, we consider the exchange-type diagram, for which we obtain
\begin{align}
    -&\frac{e^2}{2T} \int_{x,y} D_{ij}(x-y)\sum_{a,b}\text{Tr}\bigg[ \gamma^i \psi_{a}(\pmb{x})\bar{\psi}_a(\pmb{y})e^{-i \epsilon_a(x^0-y^0)} \nonumber \\
    &\times \gamma^j \psi_b(\pmb{y})\bar{\psi}_b(\pmb{x})e^{-i\epsilon_b(y^0-x^0)}\bigg] \nonumber \\
    &= -\frac{e^2}{2}\sum_{a,b}\int_{\pmb{x},\pmb{y}} \psi^{\dagger}_a(\pmb{x})\psi^{\dagger}_b(\pmb{y}) \alpha^i\alpha^j D_{ij}(\pmb{x}-\pmb{y};\epsilon_a-\epsilon_b) \nonumber \\
    &\times\psi^{}_a(\pmb{y})\psi^{}_b(\pmb{x}).
\end{align}
This retardation correction can then be written as the following energy shift
\begin{align}
    \Delta E^{(1,\text{core})}_{\text{Breit,ret}} &= \frac{1}{2}\sum_{a,b} \bra{ab}I^r(0)\ket{ab} - \bra{ab}I^r(\epsilon_a-\epsilon_b)\ket{ba}, 
\end{align}
with the transverse, retarded photon kernel $I^r(\omega)$ defined in accordance with Appendix \ref{app:mat_elem}.

\section{Expansion of the Effective Action around the no-pair DHFC Saddle}

The construction of the effective action contains the implicit assumption that the propagator $G$ on which we evaluate the functional is the solution to the DSEs derived from the full functional. Given that the $G_{\text{DHFC}}$ propagator is the solution to the DSE arising from the truncated functional $\Gamma = \Gamma_0 -i \tilde{\Phi}^{(1)}$, we choose it as an expansion point to obtain corrections to the corresponding energy order by order
\begin{align}
    \Delta\Gamma[G] &= 
    \Delta\Gamma[G_{\text{DHFC}}] + \Delta\Gamma'[G_{\text{DHFC}}](G-G_{\text{DHFC}}) \nonumber \\
    &+ \dots. \label{eq:expansion}
\end{align}
Note that the choice of expansion point for the effective action is not necessarily unique. Any non-interacting propagator, e.g., the propagator in the Kohn-Sham potential from density functional theory may be used instead~\cite{10.1063/1.1650307}. In that case, the linear term does not vanish at first order and generates higher order corrections. 
The kinetic/zeroth order contribution to the effective action is then
\begin{align}
  \Gamma_0[G_{\text{DHFC}}]  &= -i\text{Tr}\ln \Bigg[G_{\text{DHFC}}^{-1}\Bigg] -i\text{Tr}\Bigg[G_{\text{DHFC}}G_0^{-1}\Bigg].
\end{align}
The first term is given by 
\begin{align}
   i\text{Tr}\ln \Bigg[G_{\text{DHFC}}^{-1}\Bigg] &= iT \sum_n \int_{\omega}\ln (\omega-\epsilon_n + i \gamma_n)   + \text{const}
\end{align}
with an infinite, but state independent constant that drops out when considering a reference (the unoccupied atom in our case). After performing the integration, one then obtains 
\begin{align}
   -i\text{Tr}\ln \Bigg[G_{\text{DHFC}}^{-1}\Bigg] + i\text{Tr}\ln \Bigg[\tilde{G}_{\text{DHFC}}^{-1}\Bigg] & = -T\sum_{a\in \text{core}}\epsilon_a,
\end{align}
matching the expected result of the purely kinetic contribution to the ground state energy.  The second term pertaining to $\Gamma_0$ is then
\begin{align}
   i \text{Tr}\Bigg[ G_{\text{DHFC}}&(G_{\text{DHFC}}^{-1}-\tilde{\Phi}^{'(1)}[G_{\text{DHFC}}])\Bigg]  \nonumber \\ &= -i \text{Tr}\Bigg[ G_{\text{DHFC}}\tilde{\Phi}^{'(1)}[G_{\text{DHFC}}])\Bigg]+\text{const},
\end{align}
which, including the reference, turns out to yield 
\begin{align}
   T \sum_{a,b\in \text{core}} g_{abab}-g_{abba},
\end{align}
where we adopt the notation from Ref.\ \cite{Johnson2007-ip}, where $g_{abcd}$ denotes a Coulomb matrix element, 
which can be found in Appendix \ref{app:mat_elem}.
The zeroth order energy is 
\begin{align}
    -\frac{\Delta \Gamma_0-i\tilde{\Phi}^{(1)}[G^+_{\text{DHFC}}]}{T} = \sum_{a}\epsilon_a -\frac{1}{2}\sum_{ab}g_{abab}-g_{abba},
\end{align}
where we used that $i\tilde{\Phi}^{(1)}[G^+_{\text{DHFC}}] = T/2 \sum_{a,b}g_{abab}-g_{abba}$.
To recap, we now recovered the energy formula and self consistent equations for the DHFC procedure from the effective action. Additional shifts to this energy can now be computed either perturbatively, or via resummation schemes. The remaining task is essentially computing $\Phi^{(2+)}$ using an appropriate approximation.\par 

In the following sections, we will compute the second order instantaneous contributions, and show that this pertubative treatment yields the known shifts from RMBPT in addition to radiative and retardation corrections that we will cover in future work. We will further compute the first-order retardation corrections to the DHFC screening.

\section{Perturbative Treatment of the Second-Order Skeletons}

We start once again by splitting the interaction into the retarded and instantaneous part within the Coulomb gauge.
The instantaneous second-order planar (direct) diagram in the skeleton expansion is then given by the expression 
\begin{align}
    \tilde{\Phi}^{(2)}_{\text{dir}}[G] &= \frac{(-i)^2e^4}{4}\int_{x_1,y_1,x_2,y_2}V(x_1-y_1)V(x_2-y_2)\nonumber \\
    &\times \text{Tr}(\gamma^0 G(x_1,x_2)\gamma^0G(x_2,x_1))\nonumber \\
    &\times\text{Tr}(\gamma^0 G(y_1,y_2)\gamma^0G(y_2,y_1)).
\end{align}
To compute this, we need to perform complex contour integration, 
where the positions of poles are fixed by the condition that the integral does not vanish, i.e., there need to be poles in each half plane for every frequency integral. From this assignment, we know that there are $3$ contributions, resulting in 
\begin{align}
    \tilde{\Phi}^{(2)}_{\text{dir}}[G] &= \frac{iT}{2}\sum_{abmn} \frac{g_{mnab}g_{mnab}}{\epsilon_m+\epsilon_n-\epsilon_a-\epsilon_b}\nonumber \\
    &+\frac{iT}{2}\sum_{\substack{mn \\
    \epsilon_l,\epsilon_k< 0}}\frac{g_{mnlk}g_{mnlk}}{\epsilon_m+\epsilon_n-\epsilon_l-\epsilon_k}
    \nonumber \\
    &+\frac{iT}{2}\sum_{\substack{mna\\\epsilon_l <0}}\frac{g_{mnla}g_{mnla}+g_{mnal}g_{mnal}}{\epsilon_m+\epsilon_n-\epsilon_l-\epsilon_a},\nonumber \\
    \tilde{\Phi}^{(2)}_{\text{dir}}[\tilde{G}]&= \frac{iT}{2}\sum_{\substack{ab \\
    \epsilon_l,\epsilon_k< 0}}\frac{g_{ablk}g_{ablk}}{\epsilon_a+\epsilon_b-\epsilon_l-\epsilon_k}\nonumber \\
    &+\frac{iT}{2}\sum_{\substack{mn \\
    \epsilon_l,\epsilon_k< 0}}\frac{g_{mnlk}g_{mnlk}}{\epsilon_m+\epsilon_n-\epsilon_l-\epsilon_k}\nonumber \\
    &+\frac{iT}{2}\sum_{\substack{am \\
    \epsilon_l,\epsilon_k< 0}}\frac{g_{amlk}g_{amlk}+g_{malk}g_{malk}}{\epsilon_a+\epsilon_m-\epsilon_l-\epsilon_k},
\end{align}
such that the associated energy shift is given by
\begin{align}
    \Delta E^{(2)}_{\text{inst,dir}} &= -\frac{1}{2}\sum_{abmn} \frac{g_{mnab}g_{mnab}}{\epsilon_m+\epsilon_n-\epsilon_a-\epsilon_b} \nonumber \\
    &+\frac{1}{2}\sum_{\substack{ab \\
    \epsilon_l,\epsilon_k< 0}}\frac{g_{ablk}g_{ablk}}{\epsilon_a+\epsilon_b-\epsilon_l-\epsilon_k} \nonumber \\
    &- \frac{1}{2}\sum_{\substack{mna\\\epsilon_l <0}}\frac{g_{mnla}g_{mnla}+g_{mnal}g_{mnal}}{\epsilon_m+\epsilon_n-\epsilon_l-\epsilon_a} \nonumber \\
    &+\frac{1}{2}\sum_{\substack{am \\
    \epsilon_l,\epsilon_k< 0}}\frac{g_{amlk}g_{amlk}+g_{malk}g_{malk}}{\epsilon_a+\epsilon_m-\epsilon_l-\epsilon_k}.\label{eq:2nd-dir}
\end{align}
We can see that the first term is exactly the direct contribution of second order RMBPT \cite{Johnson2007-ip}, whereas the other terms including negative energy states are associated with radiative and pair corrections that we choose to bundle with the retarded interactions.
Similarly, we can evaluate the second order instantaneous exchange diagram as
\begin{align}
  \tilde{\Phi}^{(2)}_{\text{exch}}[G] &= -\frac{iT}{2}\sum_{abmn} \frac{g_{nmab}g_{mnab}}{\epsilon_m+\epsilon_n-\epsilon_a-\epsilon_b} \nonumber \\
  &-\frac{iT}{2}\sum_{\substack{mn\\\epsilon_l,\epsilon_k< 0}} \frac{g_{nmlk}g_{mnlk}}{\epsilon_m+\epsilon_n-\epsilon_l-\epsilon_k} \nonumber \\
  &-\frac{iT}{2}\sum_{\substack{mna\\\epsilon_l< 0}} \frac{g_{nmla}g_{mnla}+g_{nmal}g_{mnal}}{\epsilon_m+\epsilon_n-\epsilon_l-\epsilon_a}, \nonumber \\
  \tilde{\Phi}^{(2)}_{\text{exch}}[\tilde{G}] &= -\frac{iT}{2} \sum_{\substack{mn\\\epsilon_l,\epsilon_k< 0}} \frac{g_{nmlk}g_{mnlk}}{\epsilon_m+\epsilon_n-\epsilon_l-\epsilon_k} \nonumber \\
  &- \frac{iT}{2} \sum_{\substack{ab\\\epsilon_l,\epsilon_k< 0}} \frac{g_{ablk}g_{balk}}{\epsilon_a+\epsilon_b-\epsilon_l-\epsilon_k}\nonumber \\
  &-\frac{iT}{2}\sum_{\substack{am \\
    \epsilon_l,\epsilon_k< 0}}\frac{g_{malk}g_{amlk}+g_{amlk}g_{malk}}{\epsilon_a+\epsilon_m-\epsilon_l-\epsilon_k}.
\end{align}
Such that, analogous to the direct diagram, we obtain the related energy shift
\begin{align}
    \Delta E^{(2)}_{\text{inst,exch}} &= \frac{1}{2}\sum_{abmn}\frac{g_{nmab}g_{mnab}}{\epsilon_m+\epsilon_n-\epsilon_a-\epsilon_b} \nonumber \\
    &-\frac{1}{2} \sum_{\substack{ab\\\epsilon_l,\epsilon_k< 0}} \frac{g_{ablk}g_{balk}}{\epsilon_a+\epsilon_b-\epsilon_l-\epsilon_k}\nonumber \\
    &+\frac{1}{2}\sum_{\substack{mna\\\epsilon_l< 0}} \frac{g_{nmla}g_{mnla}+g_{nmal}g_{mnal}}{\epsilon_m+\epsilon_n-\epsilon_l-\epsilon_a}\nonumber \\
        &-\frac{1}{2}\sum_{\substack{am \\
    \epsilon_l,\epsilon_k< 0}}\frac{g_{malk}g_{amlk}+g_{amlk}g_{malk}}{\epsilon_a+\epsilon_m-\epsilon_l-\epsilon_k}\label{eq:2nd-exch}.
\end{align}
Thus, we can see that the total second order RMBPT energy shift is given by 
\begin
{align}
    \Delta E_{\text{inst,no-pair}}^{(2)} &= -\frac{\Phi^{(2)}[G^+]}{iT} \nonumber \\&= -\frac{1}{2}\sum_{abmn} \frac{g_{mnab}(g_{mnab}-g_{nmab})}{\epsilon_m+\epsilon_n-\epsilon_a-\epsilon_b},
\end{align}
matching the well-known results known from second order RMBPT \cite{Johnson2007-ip}. Second order introduces two separate demands on top of the classification itself: a combinatorial increase in the number of cutting-rule terms to evaluate, and the need for degenerate-state (quasi-degenerate) perturbation theory, which is a standard requirement of second-order many-body treatments independent of this classification.  We defer the treatment of those issues to future work.\par 

This treatment, of course, can be extended to higher orders: the skeleton expansion generates higher‑order ladder and crossed-ladder diagrams. At third order and beyond,
the RMBPT diagrams with self-energy insertions beyond first order are not skeletons;
they are recovered from the expansion of $G$ around $G_{\rm DHFC}$ in
Eq.~(\ref{eq:expansion}), so that the correspondence holds order by order. The interesting part, however, is the systematic inclusion of QED corrections into these many-body methods. The framework we used to derive these results makes it quite apparent what approximations we used to arrive at this point, meaning that a systematic treatment of omitted terms/effects becomes tractable.

\section{Atoms with one valence electron}

Similarly to the traditional way to compute RMBPT corrections, valence electrons need a separate treatment. Computing the diagrams from the expansion of the LW functional yields corrections to the systems energy due to its vacuum structure. Valence electrons do not contribute at that point, and need to be treated separately. In order to do so, we consider the self-energy diagrams associated with the two second-order contributions, i.e., 
\begin{align}
    \tilde{\Sigma}^{(2)}[G^+] &= \frac{\delta \tilde{\Phi}^{(2)}[G^+]}{\delta G^+}.
\end{align}

We further note that one diagram is topologically equivalent to an overlapping loop correction, i.e., 
\begin{align}
    \begin{gathered}
    \begin{fmffile}{SE_5}
        \begin{fmfgraph*}(70,20)
            \fmfright{i1,i2}
            \fmfleft{o1,o2}
            \fmf{phantom,tension=1.5}{i2,v21}
            \fmf{phantom,tension=1.5}{v22,o2}
            \fmf{double,tension=1.5}{i1,v11}
            \fmf{phantom,tension=.5}{v11,v12}
            \fmf{double,tension=1.5}{v12,o1}
            \fmf{double,label=$+$,label.dist=1pt,tension=.5,right=.2}{v21,v22}
            \fmffreeze
            \fmf{dashes}{v12,v22}
            \fmf{dashes}{v11,v21}   
            \fmf{double}{v11,v22}
            \fmf{double}{v21,v12}
            \fmfipair{vmid}
            \fmfiequ{vmid}{(35,10)}
            \fmfiv{label=$+$,l.a=-90,l.d=3pt}{vmid}
            \fmfiv{label=$+$,l.a=90,l.d=3pt}{vmid}
            \fmfv{decor.shape=circle,decor.filled=full,decor.size=1.4thick}{v11,v22,v21,v12}
        \end{fmfgraph*}
    \end{fmffile}
    \end{gathered} 
    = 
    \vcenter{\vbox{\kern16pt\hbox{
    \begin{fmffile}{SE_6}
        \begin{fmfgraph*}(70,30)
            \fmfleft{o1}
            \fmfright{i1}
            \fmf{double,tension=1.5}{i1,v11}
            \fmf{double,label=$+$,label.dist=1pt,tension=.5}{v11,v12}
            \fmf{double,label=$+$,label.dist=1pt,tension=1}{v12,v22}
            \fmf{double,label=$+$,label.dist=1pt,tension=.5}{v22,v23}
            \fmf{double,tension=1.5}{v23,o1}
            \fmf{dashes,tension=0,left=1}{v11,v22}
            \fmf{dashes,tension=0,left=1}{v23,v12}
            \fmfv{decor.shape=circle,decor.filled=full,decor.size=1.4thick}{v11,v22,v12,v23}
        \end{fmfgraph*}  
    \end{fmffile}}}}
\quad,
\end{align}
meaning that the different possibilities of opening one fermion line in the non planar diagram yields the two second-order self-energy insertions. 
Evaluating said diagrams on the valence state in question then results in the induced correlation shift. This is essentially the result one obtains from applying the two-time Green's function method~\cite{Shabaev2002} to the diagrams shown above, where the subtraction vanishes, since there is no first order shift given the DHFC reference potential we use. The other relevant diagram describes a part of the loop-after-loop Lamb shift contribution in which the full photon propagators are substituted with Coulomb propagators, and electron propagators are treated in the no-pair approximation.\par
It is notable that not including the valence state(s) in the vacuum is a choice we
do not \textit{have} to make. It is, however, a physically motivated choice, since
it gives direct access to the ionization energy of the valence state. Equivalently,
one may include the valence electron in the vacuum and choose the closed-shell core
as reference instead of the QED vacuum; the resulting ``ground-state energy''
$E_0=-\Delta\Gamma/T$ is then the negative ionization energy, and at second order
the core contributions cancel, reproducing Eqs.~(\ref{eq:val_dir}) and
(\ref{eq:val_exch}). With the QED vacuum as reference, by contrast, the correlation
correction of the valence electron is mixed into the core corrections and cannot be
extracted easily.
We calculate the correlation correction to the valence electron as 
\begin{align}
    \bra{v}\Sigma^{(2)}_{\text{dir}}\ket{v} &= (-i)^2e^4\int_{w,z,y_1,y_2} V(w-y_1)V(z-y_2)\nonumber \\
    &\times \text{Tr}(\gamma^0G^+(y_1,y_2)\gamma^0G^+(y_2,y_1)) \nonumber \\
    &\times \psi^{\dagger}_v(w)G^+(w,z)\gamma^0\psi_v(z),
\end{align}
where the time-dependent orbital is $\psi_v(z) = e^{-i\epsilon_vz^0}\psi_v(\pmb{z})$, where $\epsilon_v$ is the energy we obtain by iterating the DHFC equation (\ref{eq:dhf}) and solving in the resulting screened background. After the frequency integration, we obtain 
\begin{align}
    \bra{v}\Sigma^{(2)}_{\text{dir}}\ket{v} &= iT\sum_{cdn}\frac{g_{nvcd}\,g_{nvcd}}{\epsilon_c+\epsilon_d-\epsilon_n-\epsilon_v}
\nonumber \\
&+ iT\sum_{cmn}\frac{g_{nmcv}\,g_{nmcv}}{\epsilon_n+\epsilon_m-\epsilon_v-\epsilon_c}. \label{eq:val_dir}
\end{align}
The other two self energy-type diagrams give the exchange correction, given by
\begin{align}
    \bra{v} \Sigma_{\text{exch}}^{(2)}\ket{v} &=-iT\sum_{cdn} \frac{g_{nvcd}g_{nvdc}}{\epsilon_c+\epsilon_d-\epsilon_n-\epsilon_v}\nonumber \\
    &-iT\sum_{cmn}\frac{g_{nmcv}g_{nmvc}
    }{\epsilon_n+\epsilon_m-\epsilon_v-\epsilon_{c}}.\label{eq:val_exch}
\end{align}
The corresponding energy shift is $\Delta E^{(2)}_v = \frac{i}{T}\bra{v}\Sigma^{(2)}\ket{v}$.
With these contributions, the total shift matches the formulas known in the literature (see e.g. \cite{Johnson2007-ip}).\par
In a similar fashion, one easily obtains the frequency-dependent Breit corrections for the valence electron, that is, 
\begin{align}
    \Delta E^{(1,\text{valence})}_{\text{Breit,ret}} = -\sum_a \bra{va}I^r(\epsilon_v-\epsilon_a)\ket{av}.
\end{align}
Similarly to the previous section, one can obtain the retardation corrections at this point, which we once again defer to future dedicated work on this issue.

\section{Beyond the No-Pair Approximation}

We saw in previous sections that the LW functional provides a natural organizational principle for many-body contributions to the ground-state energy of atomic systems. We obtained the known no-pair RMBPT corrections, along with a number of radiative and pair corrections. Not all of those corrections, however, are to be treated on the same footing: Radiative corrections generically already appear for hydrogen-like atoms, and need to be renormalized. Here, we treat pair-corrections, which are given by the second terms in Eqs.\ \eqref{eq:2nd-dir}, \eqref{eq:2nd-exch} 
\begin{align}
    \Delta E_{\text{inst,pair}}^{(2)}
    = -\frac{1}{2}\sum_{ab}\sum_{\substack{l,k\\ \epsilon_l,\epsilon_k<0}}
      \frac{g_{lkab}\,(g_{lkab}-g_{klab})}{\epsilon_l+\epsilon_k-\epsilon_a-\epsilon_b},
\end{align}
where we take the negative energy states to be solutions of the Dirac-Coulomb equations within the converged DHFC potential, as implied by the expansion point of the effective action. We give the numerical values for this interaction for a number of elements in the following section, but note that this may be a larger effect for retarded Breit contributions (see, e.g., Ref.\ \cite{PhysRevA.51.1167}) which we cover in aforementioned future work. \par
As a small aside, we can estimate the number of partial waves needed for the numerical evaluation of the term shown above. We assume a semiclassical expression for the angular momentum, and use that the relevant negative-energy states are nearly free with momenta
$p \sim m_e$: the coupling to the bound electrons is suppressed as $p/m_e$ for
$p \ll m_e$, while the energy denominator suppresses $p \gg m_e$. The angular momentum needed for correlation with a state at radius $r\sim a_0/Z$ has then to be larger than\footnote{This is an estimate of when convergence \textit{starts}.} 
\begin{align}
    l \gtrsim pr \sim (Z \alpha)^{-1}.
\end{align}
It is thus clear that computing this correction is most expensive for light systems, for which the effect is smallest.

\section{Exemplary Computations}

\begin{table}[thb]
\begin{ruledtabular}
\begin{tabular}{c l l l l}
   $Z$  & $ E_{\text{DHFC}}$& $\epsilon_{3s}$ &$\Delta E^{(2)}_{\text{core}}$ & $\Delta E^{(2)}_{\text{val}}$\\ \hline\rule{0pt}{1\normalbaselineskip}
   $26$  & $1148.42596$ & $17.96351$ & $0.40795$ & $0.025847$ \\
         &          &           &       &      $0.025 852$\footnotemark[2]     \\
   $30$  &$1568.58062$& $27.08374$& $0.41163$ & $0.02715$\\
         &          &    $27.08455$\footnotemark[1]       &       &           \\
   $35$ & $2189.38029$ & $41.12452$&$0.41560$ & $0.02840$ \\
   $40$ & $2918.74449$ & $58.16216$ & $0.41926$ & $0.02941$ \\
   $45$ & $3759.68455$ &$78.28043$ & $0.42289$ & $0.03029$ \\
   $50$ & $4715.83170$ & $101.58190$ & $0.42666$ & $0.03112$ \\
   $55$ & $5791.51942$ & $128.19018$ & $0.43073$ & $0.03193$
\end{tabular}
\end{ruledtabular}
\footnotetext[1]{$E_0+E_1$ in the Kohn--Sham potential, Ref.~\cite{PhysRevA.91.062508};
equal to the DHF eigenvalue up to second-order terms in $V_{\rm HF}-V_{\rm KS}$.}
\footnotetext[2]{Ref.\ \cite{PhysRevA.38.2699}}
\caption{Second-order RMBPT shifts (fourth and fifth columns) along with the converged DHFC energies (second column) for the ground state energy of sodiumlike ions in the $Z=26-55$ region. Nuclear radii were taken from Ref.\ \cite{Angeli2013} and we assumed a two-parameter Fermi nuclear charge distribution for all elements. All results are given in atomic units.}
\label{tab:rmbpt}
\end{table}

\begin{table}[thb]
\begin{ruledtabular}
\begin{tabular}{c l l l}
   $Z$  & $ E_{\text{DHFC}}$& $\Delta E^{(1,\text{core})}_{\text{Breit,ret}}$ & $\Delta E^{(1,\text{valence})}_{\text{Breit,ret}}$\\ \hline\rule{0pt}{1\normalbaselineskip}
   $26$ &  $1148.42596$  &$-0.43580$  & $-0.005480$\\
        &   & &$-0.005480$\footnotemark[1] \\
   $30$  &$1568.58062$& $-0.69492$ & $-0.00951$\\
   $35$ & $2189.38029$ & $-1.14409$ & $-0.01684$\\
   $40$ & $2918.74449$ & $-1.75761$ & $-0.02727$\\
   $45$ & $3759.68455$ &$-2.56365$  & $-0.04144$\\
   $50$ & $4715.83170$ & $-3.59198$ & $-0.06006$\\
   $55$ & $5791.51942$ & $-4.87442$ & $-0.08390$    
\end{tabular}
\end{ruledtabular}
\footnotetext[1]{Ref.\ \cite{PhysRevA.38.2699}}
\caption{Frequency-dependent Breit corrections to the ground state energy of sodiumlike ions in the $Z=26-55$ region.}
\label{tab:breit_freq}
\end{table}

The above expressions contain matrix elements of Dirac orbitals. We give a few exemplary numerical evaluations in terms of binding energies, that is, the negatives of the previously defined energies and energy shifts. In our numerical evaluations, the radial functions are represented as basis sets constructed from B-splines~\cite{Johnson2007-ip}, based on the implementation of Ref.~\cite{Cakir2020}. The large and small radial functions are treated within the dual kinetic balance approach~\cite{Shabaev2004} to avoid spurious solutions. The reduced Coulomb and photon exchange matrix elements are given in Appendix~\ref{app:mat_elem}. Radial integrals were evaluated with Gauss-Legendre quadrature. For the numerical solution of the Dirac-Hartree-Fock equations~\eqref{eq:dhf}, we adopted the method of Ref.~\cite{Johnson2007-ip} to our B-spline representation, and additionally employed Pulay's method to accelerate the self-consistent iterations~\cite{Pulay1980} (cf.\ Appendix \ref{app:pulay} for details).

First, we give the second-order RMBPT+DHFC results for energy shifts for sodiumlike ions in Tab.\ \ref{tab:rmbpt}. We further give the first-order frequency-dependent Breit corrections in Tab.~\ref{tab:breit_freq}, and the second-order pair corrections for He-like ions in Tab.~\ref{tab:pair-corrections}, which we compare with the QED results of Ref.~\cite{PhysRevA.51.1167}. Our results in Tabs.~\ref{tab:rmbpt}, \ref{tab:breit_freq} match the ones found in the literature. This agreement reflects the fact that we rederived exactly the known formulae underlying existing literature, and serves in part as a test of our numerical implementation. The pair corrections in Tab.~\ref{tab:pair-corrections} agree with the Furry-picture QED results of Ref.~\cite{PhysRevA.51.1167}, with a difference decreasing from $6\%$ at $Z=30$ to $0.15\%$ at $Z=92$. This is consistent with the use of point-like nuclei and bare Coulomb orbitals in Ref.~\cite{PhysRevA.51.1167}, as opposed to finite nuclei and DHFC orbitals in this work. After this verification step, we present pair corrections in Ne-like ions in Tab.~\ref{tab:ne-pair}. In case of heavy ions of this isoelectronic series, virtual pair corrections must be taken into account to reach mHa accuracy.

\begin{table}[thb]
    \centering
    \begin{ruledtabular}
    \begin{tabular}{c l l}
       $Z$ & This work & Ref.\ \cite{PhysRevA.51.1167}  \\
        \hline\rule{0pt}{1\normalbaselineskip}
         $30$& $-3.836(65)\times10^{-5}$& $-4.07\times10^{-5}$ \\
         $40$ & $-9.501(80)\times10^{-5}$ & \\
         $50$ & $-1.929(10)\times10^{-4}$ & \\
         $60$ & $-3.474(12)\times10^{-4}$ & $-3.56\times10^{-4}$\\
         $70$ & $-5.778(14)\times10^{-4}$ & $-5.90\times10^{-4}$ \\
         $92$ & $-1.5009(21)\times10^{-3}$ & $-1.5031\times10^{-3}$
    \end{tabular}
    \end{ruledtabular}
    \caption{Pair corrections to second order RMBPT for He-like ions. We compare with Ref.\ \cite{PhysRevA.51.1167}, in which the authors extract this pair contribution to our quantity from two-photon exchange in the Coulomb gauge\footnote{The reference uses pointlike nuclei and bare Coulomb potentials, therefore, we do not expect the two values to agree exactly.}. The quoted uncertainties are estimated from the extrapolation of the partial wave expansion beyond $l=30$. All results are given in atomic units.}
    \label{tab:pair-corrections}
\end{table}

\begin{table}[thb]
    \centering
    \begin{ruledtabular}
    \begin{tabular}{c l l}
         $Z$ & $\Delta E^{(2)}_{\text{no-pair}}$ & $\Delta E^{(2)}_{\text{pair}}$ \\
        \hline\rule{0pt}{1\normalbaselineskip}
         $30$ & $4.117340(92)\times10^{-1}$ & $-5.48(27)\times10^{-5}$ \\
         $40$ & $4.193780(74)\times10^{-1}$ & $-1.451(47)\times10^{-4}$ \\
         $50$ & $4.267848(54)\times10^{-1}$ & $-3.108(67)\times10^{-4}$ \\
         $60$ & $4.353699(34)\times10^{-1}$ & $-5.865(86)\times10^{-4}$ \\
         $70$ & $4.462273(13)\times10^{-1}$ & $-1.020(11)\times10^{-3}$ \\
         $80$ & $4.606438(13)\times10^{-1}$ & $-1.681(13)\times10^{-3}$ \\
         $92$ & $4.855369(43)\times10^{-1}$ & $-2.928(15)\times10^{-3}$
    \end{tabular}
    \end{ruledtabular}
    
    \caption{Pair corrections to second order RMBPT for Ne-like ions. The quoted uncertainties are estimated from the extrapolation of the partial wave expansion beyond $l=30$. All results are given in atomic units.}
    \label{tab:ne-pair}
\end{table}

\section{Conclusion and Outlook}

In this work, we presented a derivation of the known RMBPT formulae for Coulomb- and
Breit-induced energy shifts in many-electron atoms and ions. In contrast to existing treatments, we obtained these results from the 2PI effective action of QED, making the omitted diagrams directly apparent. Those include radiative corrections, as well as retardation and pair corrections to the known RMBPT shifts. As an example, we evaluated the virtual electron-positron pair corrections to second-order RMBPT, which connect to Furry-picture QED results for He-like ions, and provided them for Ne-like ions. The same procedure extends to transverse photon exchange at second order, which enables a consistent treatment of frequency-dependent Breit corrections at that order.
The diagrammatic interpretation of RMBPT shown in this work may be extended by identifying
existing many-body methods as partial (re)summations of classes of skeletons. Moreover, choosing a local expansion point such as a Kohn-Sham propagator places correlation and radiative corrections in the same extended Furry picture, allowing the use of methods developed for that picture.

\begin{acknowledgments}

We thank Halil Cakir for making his Fortran 90 B-spline code
base available to us, and Jonas Sommerfeldt for helpful discussions. We also thank Konstantin Beyer, Viktoria Noel, Chunhai Lyu, and Matteo Moretti for comments on the manuscript.
This work is supported by the Deutsche Forschungsgemeinschaft (DFG, German Research Foundation) Project-ID 273811115—SFB 1225, and comprises parts of the PhD thesis work of
C.\ d.\ J.\ to be submitted to Heidelberg University.
\end{acknowledgments}

\bibliographystyle{apsrev4-1}
\bibliography{references}

\appendix

\section{Matrix Elements}\label{app:mat_elem}
The Coulomb interaction matrix element is given by the general expression \cite{Johnson2007-ip}
\begin{align}
    g_{abcd} = \int_{\pmb{x},\pmb{y}} \psi^{\dagger}_a(\pmb{x})\psi_c(\pmb{x})V(\pmb{x}-\pmb{y})\psi_{b}^{\dagger}(\pmb{y})\psi_d(\pmb{y}).
\end{align}
Plugging in the multipole expansion for the Coulomb interaction, and performing angular integrations, we obtain
\begin{align}
    g_{abcd} = \sum_{L=0}^{\infty}R^c_L(abcd)J_L(abcd),
\end{align}
with the angular function $J_L$ defined as
\begin{align}
    J_{L}(abcd) &= \sum_{m_L=-L}^{L} \Bigg[(-1)^{j_a-m_a+j_b-m_b}\nonumber \\[.5em] &\times 
    \begin{pmatrix}
        j_a & L & j_c\\
        -m_a& m_L & m_c
    \end{pmatrix}
    \begin{pmatrix}
        j_b & L & j_d\\
        -m_b& -m_L & m_d
    \end{pmatrix}\Bigg],\nonumber \\
\end{align}
and radial integral 
\begin{align}
    R^c_L(abcd) &= (-1)^L \langle \kappa_a ||C_L||\kappa_c\rangle \langle \kappa_b ||C_{-L}||\kappa_d\rangle \nonumber \\
    &\times \int_{r_1,r_2} W_{ac}(r_1)W_{bd}(r_2)\frac{r_<^L}{r_>^{L+1}},
\end{align}
where we defined the following combinations of radial wavefunctions for brevity
\begin{align}
    W_{ab}(r) &= F_a(r)F_b(r)+G_a(r)G_b(r)\nonumber \\[.5em]
    V_{ab}(r) &= -G_a(r)F_b(r) - F_a(r)G_b(r) \nonumber \\[.5em]
    U_{ab}(r) &= -G_a(r)F_b(r) +F_a(r)G_b(r),
\end{align}
of which we will use the latter two in the following. 
The photon exchange matrix element can be decomposed in a similar fashion, where we have 
\begin{align}
    \bra{ab}I(\omega)\ket{cd} = \sum_{L=0}^{\infty} J_L(abcd)R_L(abcd;\omega),  
\end{align}
where the radial integral has a somewhat more complicated structure that is gauge dependent. In the following, we give the expressions for the matrix element in Coulomb gauge, Feynman gauge expressions for the same matrix element can be found in Ref.\ \cite{PhysRevA.46.3762}. We write 
\begin{align}
    R_L(abcd;\omega) = R_{L,1}(abcd;\omega) + R_{L,2}(abcd;\omega),
\end{align}
where 
\begin{align}
    R_{L,1}(abcd;\omega) &=  (-1)^L \langle \kappa_a ||C_L||\kappa_c\rangle \langle \kappa_b ||C_{-L}||\kappa_d\rangle \nonumber \\[.25em]
    &\times \mathcal{I}_L(abcd;\omega) \nonumber \\[.5em]
    R_{L,2}(abcd;\omega)  &= (-1)^{L+1} \langle -\kappa_a ||C_L||\kappa_c\rangle \langle -\kappa_b ||C_{-L}||\kappa_d\rangle \nonumber \\[.25em]
    &\times \frac{(\kappa_a+\kappa_c)(\kappa_b+\kappa_d)}{L(L+1)}\mathcal{J}_L(abcd;\omega).
\end{align}
Here, we used two different radial integrals. The first one being 
\begin{align}
    \mathcal{J}_L(abcd;\omega) &= i\omega(2L+1)\nonumber \\
    &\times \int_{r_1,r_2} j_L(\omega r_<)h^{(1)}_L(\omega r_>)V_{ac}(r_1)V_{bd}(r_2),
\end{align}
where $j_L$ and $h^{(1)}_L$ are the spherical Bessel function and the spherical Hankel function of the first kind, both of order $L$. The second integral is the gauge dependent part, and in the Coulomb gauge, it decomposes as
\begin{align}
    \mathcal{I}_L(abcd;\omega) &= \mathcal{I}_{L,1}(abcd;0) + \frac{L(L+1)}{2L+1}\Bigg(\frac{\mathcal{I}_{L,2}(abcd;\omega)}{2L+3}\nonumber \\
    &+ \frac{\mathcal{I}_{L,3}(abcd;\omega)}{2L-1}-\mathcal{I}_{L,4}(abcd;\omega)\nonumber \\
    &-\mathcal{I}_{L,4}(badc;\omega)\Bigg),
\end{align}
where we have introduced the expressions
\begin{align}
    \mathcal{I}_{L,1}(abcd;\omega) &= i\omega(2L+1)\nonumber \\
    &\times \int_{r_1,r_2} j_L(\omega r_<)h^{(1)}_L(\omega r_>)W_{ac}(r_1)W_{bd}(r_2) \nonumber \\
    \mathcal{I}_{L,2}(abcd;\omega) &= i\omega(2L+3)\nonumber \\
    \times \int_{r_1,r_2} &j_{L+1}(\omega r_<)h^{(1)}_{L+1}(\omega r_>)Q_{L,ac}(r_1)Q_{L,bd}(r_2) \nonumber \\
    \mathcal{I}_{L,3}(abcd;\omega) &= i\omega(2L-1)\nonumber \\
    \times \int_{r_1,r_2} &j_{L-1}(\omega r_<)h^{(1)}_{L-1}(\omega r_>)P_{L,ac}(r_1)P_{L,bd}(r_2), 
\end{align}
and the Coulomb specific integral $\mathcal{I}_{L,4}$ being given by
\begin{widetext}
    \begin{align}
         \mathcal{I}_{L,4}(abcd;\omega) 
    &= \int_{0}^{\infty} \text{d}r_1\int_{0}^{r_1}\text{d}r_2 \,\Bigg[ i\omega j_{L-1}(\omega r_2)h^{(1)}_{L+1}(\omega r_1) - \frac{2L+1}{\omega^2}\frac{r_2^{L-1}}{r_1^{L+2}}\Bigg]Q_{L,ac}(r_1)P_{L,bd}(r_2) \nonumber \\
    &+i\omega \int_{0}^{\infty}\text{d}r_1 \int_{r_1}^{\infty}\text{d}r_2 \,j_{L+1}(\omega r_1)h^{(1)}_{L-1}(\omega r_2)Q_{L,ac}(r_1)P_{L,bd}(r_2).
    \end{align}
\end{widetext}
with the combinations of radial wavefunctions 
\begin{align}
    Q_{L,ab}(r) = \frac{\kappa_b-\kappa_a}{L+1}V_{ab}(r)-U_{ab}(r)\nonumber \\[.5em]
    P_{L,ab}(r) = \frac{\kappa_b-\kappa_a}{L}V_{ab}(r) + U_{ab}(r).
\end{align}
With these results, we then simply define the transverse, retarded Breit matrix element as 
\begin{align}
     R_L^r(abcd;\omega)  = R_L(abcd;\omega)-R_L^c(abcd).
\end{align}
Using the above expressions, one can then obtain Coulomb and photon exchange corrections to energy levels given the corresponding wavefunctions and expressions in the main text.

\section{Pulay's Method}\label{app:pulay}

To speed up the convergence of the DHFC resummation, we implement Pulay's method, more commonly known as direct inversion in the iterative subspace (DIIS). This method is commonly used in computational chemistry to accelerate the convergence of self consistent field methods. This method is best understood when recasting the problem in matrix form. \par
The self consistent solution of Eq.\ (\ref{eq:dhf}) is an iterative fixed-point problem, that is we iterate 
\begin{align}
    h_D\cdot\psi_i^{(n)} + \Sigma^{(n)}[\psi_i^{(n)}] = \epsilon^{(n)}_i\psi^{(n)}_i,
\end{align}
where the superscript $n$ indicates the order of iteration. The $n$-th order self energy operator applied to $\psi_i^{(n)}$ is given by 
\begin{align}
   \Sigma^{(n)}[\psi_i^{(n)}] &= e^2\sum_a\Bigg[\int_{\pmb{z}} \rho^{(n-1)}_{aa}(\pmb{z})V(\pmb{x}-\pmb{z})\psi_i^{(n)}(\pmb{x})\nonumber \\
   &- \int_{\boldsymbol y} V(\boldsymbol x-\boldsymbol y)\,\psi_a^{(n-1)}(\boldsymbol x)\,\psi_a^{\dagger(n-1)}(\boldsymbol y)\,\psi_i^{(n)}(\boldsymbol y)\Bigg].
\end{align}
This can be recast into matrix form by projecting the above operators on the chosen basis of wavefunctions. The problem is then recast to 
\begin{align}
        F^{(n)}\pmb{c}_i = \epsilon^{(n)}_i S\pmb{c}_i,
\end{align}
where the vector $\pmb{c}_i$ contains the expansion coefficients of the wavefunctions in the chosen basis, i.e., $\psi_i = \sum_n c_i^n f_n$. Convergence of this problem can be improved via DIIS by retaining a small history of matrices $F$ for $m$ iterations to compute a deviation from self consistency, i.e., 
\begin{align}
    (\pmb{e}^{(n)})_{p+\text{dim}\cdot q} = (F^{(n)}-F^{(n-1)})_{pq},
\end{align}
where we defined the matrix $F$ to have dimension $\text{dim}\times \text{dim}$. For the iteration $n$, we then define the DIIS matrix 
\begin{align}
    F^{(n)}_{\text{DIIS}} = \sum_{i=n-m+1}^n c_i F^{(i)},
\end{align}
where the coefficients $c_i$ are chosen to minimize $|\sum_i c_i \pmb{e}^{(i)}|$ subject to the constraint $\sum_i c_i = 1$. Using a Lagrange multiplier $\lambda$, one can recast to the matrix equation 
\begin{widetext}
\begin{align}
    \begin{pmatrix}
        B_{n-m+1,n-m+1} & \cdots & B_{n-m+1,n} & -1 \\
        \vdots & \ddots & \vdots & \vdots \\
        B_{n,n-m+1} & \cdots & B_{n,n} & -1 \\
        -1 & \cdots & -1 & 0
    \end{pmatrix}
    \begin{pmatrix}
        c_{n-m+1} \\ \vdots \\ c_n \\ \lambda
    \end{pmatrix}
    =
    \begin{pmatrix}
        0 \\ \vdots \\ 0 \\ -1
    \end{pmatrix},
\end{align}
\end{widetext}
where the entries are defined as $B_{ij} = \pmb{e}_i \cdot \pmb{e}_j$. Using this, one can build an optimized matrix $F_{\text{DIIS}}$ at every order. In practice, we retain $6$ past iterations of $F$ to build the optimized matrix, and start implementing the DIIS procedure after three regular DHFC iterations. The latter is done because the extrapolation relies on the
self-consistency map $F^{(n-1)}\mapsto F^{(n)}$ being approximately affine near the fixed point, so that a linear combination of past $F^{(i)}$ approximately
inherits the same combination of their observed errors $\mathbf{e}^{(i)}$. Far from convergence, this approximation fails, which is why we perform a few plain iterations before implementing DIIS.

\end{document}